\documentclass[11pt]{article}

\PassOptionsToPackage{table}{xcolor}
\usepackage[preprint]{acl}

\usepackage{times}
\usepackage{latexsym}
\usepackage{booktabs}
\usepackage{arydshln}
\usepackage{amsmath}
\usepackage{amsfonts}
\usepackage{xspace}
\usepackage{subcaption}
\usepackage{algorithm}
\usepackage{pifont}

\usepackage{natbib}

\usepackage[T1]{fontenc}
\usepackage[utf8]{inputenc}
\usepackage{microtype}
\usepackage{inconsolata}
\usepackage{graphicx}

\usepackage{amssymb,amsthm}

\usepackage{algorithmic}
\usepackage{textcomp}
\usepackage{multirow}
\usepackage{url}
\usepackage{enumitem}
\usepackage{hyperref}
\usepackage{array}
\usepackage{tikz}
\usepackage{pgfplots}
\pgfplotsset{compat=1.18}
\usepackage{makecell}

\title{Learning-Based Automated Adversarial Red-Teaming for Robustness Evaluation of Large Language Models}
\author{
  \textbf{Zhang Wei\textsuperscript{1}},
  \textbf{Hanxuan Chen\textsuperscript{2}},
  \textbf{Peilu Hu\textsuperscript{1}},
  \textbf{Zhenyuan Wei\textsuperscript{3}},
  \textbf{Chenwei Liang\textsuperscript{3}},
  \textbf{Jiayi Gu\textsuperscript{4}},
  \textbf{Wenqian Weng\textsuperscript{5}},\\
  \textbf{Jacqueline Pang\textsuperscript{6}},
  \textbf{Hao Yan\textsuperscript{7}},
  \textbf{Li Mei\textsuperscript{1}},
  \textbf{Shengning Lang\textsuperscript{7}},
  \textbf{Kuan Lu\textsuperscript{8}},
  \textbf{Xi Xiao\textsuperscript{9}},\\
  \textbf{Zhimo Han\textsuperscript{10}},
  \textbf{Yijin Wang\textsuperscript{11}},
  \textbf{Yichao Zhang\textsuperscript{12}},
  \textbf{Chen Yang\textsuperscript{13}},
  \textbf{Zhenyu Yu\textsuperscript{14}},\\
  \textbf{Riyang Bao\textsuperscript{15}},
  \textbf{Xinyuan Song\textsuperscript{15}},
  \textbf{Junfeng Hao\textsuperscript{16}},
  \textbf{Mu-Jiang-Shan Wang\textsuperscript{3}$^\dagger$}
\\
  \textsuperscript{1}Independent Researcher \quad
  \textsuperscript{2}Hunan University \quad
  \textsuperscript{3}Shenzhen Kaihong Digital Industry Development Co., Ltd.
\\
  \textsuperscript{4}Central University of Finance and Economics \quad
  \textsuperscript{5}Wayne State University\\
  \textsuperscript{6}Cornell University \quad
  \textsuperscript{7}Stevens Institute of Technology \quad
  \textsuperscript{8}Cornell University \\
  \textsuperscript{9}Oak Ridge National Laboratory \quad
  \textsuperscript{10}Zhengzhou University of Light Industry \quad
  \textsuperscript{11}Xidian University \\
  \textsuperscript{12}The University of Texas at Dallas \quad
  \textsuperscript{13}AI Safety Research Lab, Institute of Advanced Computing \\
  \textsuperscript{14}University of Malaya \quad
  \textsuperscript{15}Emory University \\
  \textsuperscript{16}Department of Nephrology, Affiliated Hospital of Guangdong Medical University \\
  \small{
    \textbf{$\dagger$ Corresponding author:}
    \href{mailto:mjs.wang@siat.ac.cn}{mjs.wang@siat.ac.cn}
  }
}

\begin{document}
\maketitle
\begin{abstract}
Red-teaming is becoming a central part of large language model (LLM) safety evaluation, yet current practice still relies heavily on expert-written prompts or fixed benchmark suites. This creates a gap between what is easy to test and what deployed models can actually do: failures may be rare, context-sensitive, and distributed across many threat categories. We study automated red-teaming as a constrained adversarial search problem and introduce a learning-driven framework that couples category-aware attack generation with hierarchical vulnerability detection. The method starts from curated safety seeds, expands them through meta-prompt-guided and evolutionary search, and scores the resulting prompt--response pairs with lexical, semantic, and behavioral detectors. Across six threat categories on GPT-OSS-20B, the framework discovers 47 validated vulnerabilities, including 21 high-severity cases and 12 novel attack patterns. Under matched query budgets, it achieves a 3.9$\times$ higher discovery rate than manual expert red-teaming while maintaining 89\% detection accuracy and full category coverage. Ablations show that the gains do not come from more prompts alone: diversity constraints prevent template collapse, coverage constraints prevent category blind spots, and semantic detection recovers failures missed by lexical rules. These results suggest that red-teaming can be made more scalable and reproducible when treated as adaptive search rather than as a static checklist.
\end{abstract}

\begin{figure*}
    \centering
    \includegraphics[width=0.95\linewidth]{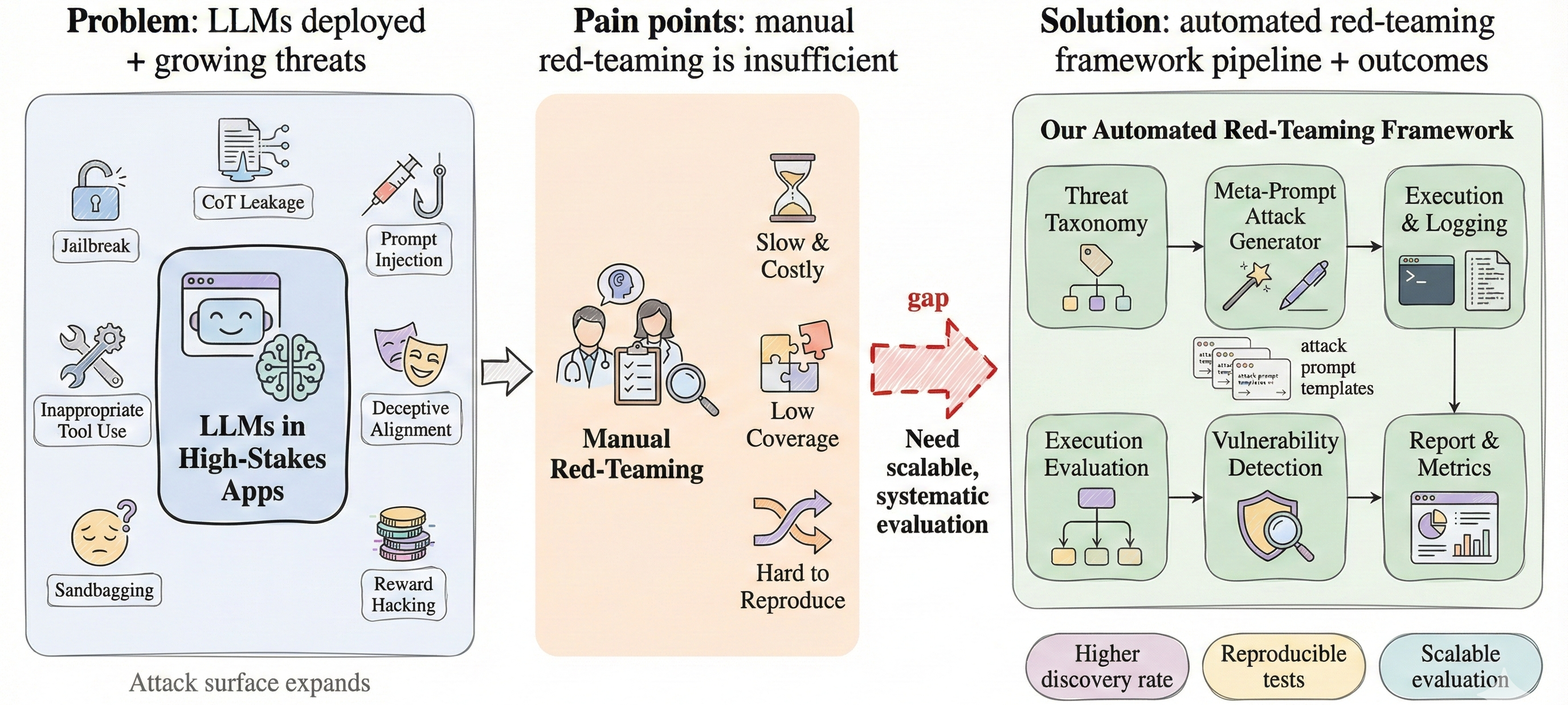}
    \caption{Why red-teaming needs adaptive search. Manual red-teaming can uncover high-value failures, but it explores only a small and difficult-to-reproduce slice of the prompt space. Fixed benchmarks improve standardization but may miss emergent or model-specific failures. Our framework fills the gap by generating diverse adversarial prompts, executing them against the target model, and scoring the resulting behaviors under a shared vulnerability rubric.}
    \label{fig:motivation}
\end{figure*}

\section{Introduction}
Large language models (LLMs) are increasingly deployed through both API-based systems and open-weight releases, from GPT-4, Claude, and Gemini to Llama and GPT-OSS models~\cite{achiam2023gpt,anthropic2024claude3,gemini2023technical,touvron2023llama,openai2025gptoss}. Their usefulness comes from flexible natural-language interaction, but the same flexibility makes safety evaluation difficult. A model can behave correctly on a fixed benchmark and still fail under a slightly different framing, a multi-turn interaction, or a tool-use scenario. Documented failures include jailbreaks, prompt injection, harmful instruction following, data leakage, reward hacking, and unsafe agent behavior~\cite{wei2023jailbroken,zou2023universal,liu2023promptinjection,carlini2021extracting,perez2022ignore,ruan2023toolemu,andriushchenko2024agentharm}.

The evaluation community has made important progress toward standardization. HarmBench, JailbreakBench, XSTest, CyberSecEval, and related safety benchmarks provide shared tasks, metrics, and failure categories~\cite{mazeika2024harmbench,chao2024jailbreakbench,rottger2024xstest,bhatt2023cyberseceval,bhatt2024cyberseceval2}. These resources are essential, but they are necessarily finite. They capture known behaviors at the time of construction, while deployed models, policies, and attack strategies keep changing. Manual red-teaming complements fixed suites by allowing expert search, yet it is expensive, hard to reproduce, and difficult to scale across model versions and threat categories~\cite{perez2022redteaming,openai2023gpt4,liang2022helm}. The central problem is therefore not whether red-teaming is useful, but how to make it systematic enough to compare models and adaptive enough to find failures that static tests miss.

Recent automated attacks show that LLMs can help generate adversarial prompts. Methods such as AutoDAN, PAIR, TAP, AdvPrompter, and simple adaptive attacks have substantially expanded the toolkit for jailbreak discovery~\cite{liu2023autodan,chao2023pair,mehrotra2023tap,paulus2024advprompter,andriushchenko2024simple}. However, many of these methods optimize for attack success in a narrow setting. A practical red-teaming system must do more: it should cover multiple threat categories, avoid collapsing onto one family of prompts, detect subtle failures rather than only explicit policy violations, and produce results that auditors can interpret and reproduce.

We address this gap by formulating automated red-teaming as a constrained adversarial search problem. The proposed framework starts with curated seeds spanning six vulnerability categories, expands them through meta-prompt-guided generation and evolutionary mutation, and evaluates each prompt--response pair with a hierarchical detector that combines lexical rules, semantic matching, and behavioral signals. Diversity and coverage constraints keep the search from degenerating into repetitive jailbreak variants, while the scoring protocol ranks findings by severity, breadth, novelty, and reproducibility.

Figure~\ref{fig:motivation} summarizes the motivation: manual red-teaming offers depth, fixed benchmarks offer comparability, and adaptive search is needed to connect the two under a reproducible evaluation budget.

This framing turns the experimental story from a leaderboard into a diagnostic question: which components make automated red-teaming discover more, cover more categories, and remain reliable under a fixed query budget? Our experiments on GPT-OSS-20B show that the full framework identifies 47 validated vulnerabilities across all six categories, compared with 12 for manual expert red-teaming, 18 for template attacks, and 23 for AdvPrompter under the same evaluation budget. The result is not merely a higher count. The framework also finds more high-severity cases, maintains 89\% detection accuracy, and continues discovering novel vulnerabilities after baseline methods plateau.

Our main contributions are:
\begin{itemize}[left = 0em]
\item We formulate automated red-teaming for LLMs as a constrained adversarial search problem with explicit diversity, coverage, and budget constraints.
\item We introduce a learning-driven framework that combines seed curation, meta-prompt-guided generation, evolutionary mutation, and hierarchical vulnerability detection across six threat categories.
\item We provide a systematic evaluation on GPT-OSS-20B showing higher discovery rate, broader category coverage, stronger novelty, and better reproducibility than manual, random, template-based, and automated baselines.
\end{itemize}

\begin{figure*}
    \centering
    \includegraphics[width=0.95\linewidth]{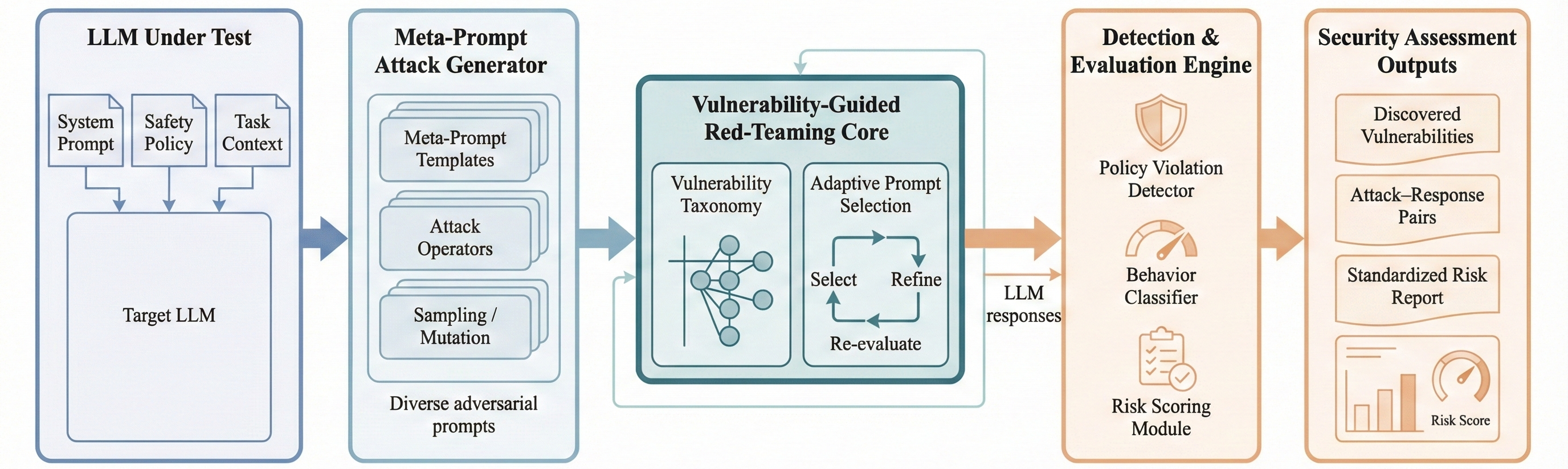}
    \caption{Overview of the proposed red-teaming framework. Curated seed prompts initialize category coverage, meta-prompt-guided generation and evolutionary mutation expand the adversarial set, and the target model is queried under a fixed budget. A hierarchical detector scores each prompt--response pair with lexical, semantic, and behavioral signals, producing a structured report over severity, breadth, novelty, and reproducibility.}

    \label{fig:overview}
\end{figure*}


\section{Related Work}

The security evaluation of LLMs sits at the intersection of adversarial machine learning, software testing, red-teaming, and benchmark design. We organize prior work around the pieces needed for a practical red-teaming system: where adversarial evaluation comes from, how LLM-specific attacks differ, what current benchmarks measure, and why adaptive search remains necessary.

\subsection{Traditional AI Safety and Adversarial Attacks}

Adversarial robustness research originally focused on closed-set prediction, where small perturbations can reliably induce misclassification~\cite{szegedy2013intriguing,goodfellow2014explaining}. Subsequent work introduced stronger attacks and optimization-based defenses~\cite{carlini2017towards,madry2017towards}, while machine-learning security surveys clarified how threat models, attacker knowledge, and evaluation protocols shape robustness claims~\cite{papernot2016towards}. These foundations are important, but LLM red-teaming differs in two ways. First, attacks are expressed in natural language rather than continuous perturbations. Second, the failure criterion is behavioral and contextual: a response may be unsafe because it leaks information, obeys an injected instruction, misuses a tool, or strategically underperforms.

\subsection{LLM-Specific Security Threats and Red-Teaming}

LLM-specific security work has documented jailbreaks, prompt injection, harmful content generation, privacy leakage, and tool-use risks~\cite{achiam2023gpt,wei2023jailbroken,liu2023promptinjection,carlini2021extracting,ruan2023toolemu}. Universal suffix attacks show that adversarial strings can transfer across aligned models~\cite{zou2023universal}, and in-the-wild studies show that jailbreak prompts arise in realistic user behavior rather than only in laboratory settings~\cite{shen2024anything,jiang2024wildteaming}. Agentic systems expand the attack surface further because failures can involve external tools, APIs, or multi-step plans~\cite{andriushchenko2024agentharm,bhatt2024cyberseceval2}. These studies motivate red-teaming beyond one-shot refusal testing: an evaluator must probe the interaction between prompt context, model policy, reasoning behavior, and deployment affordances.

\subsection{Automated Testing and Vulnerability Discovery}

Automation has long been central to software security through fuzzing and coverage-guided testing~\cite{miller1990empirical,godefroid2008automated}. LLM red-teaming has begun to adopt a similar spirit. Model-generated red-team prompts can scale behavioral testing~\cite{perez2022redteaming,perez2023discovering}, while automated jailbreak methods such as AutoDAN, PAIR, TAP, AdvPrompter, and simple adaptive attacks improve attack success under different access assumptions~\cite{liu2023autodan,chao2023pair,mehrotra2023tap,paulus2024advprompter,andriushchenko2024simple}. Our work is complementary: instead of optimizing only for a successful jailbreak, we optimize for discovery under coverage, diversity, novelty, and reproducibility constraints across multiple vulnerability categories.

\subsection{AI Alignment and Safety Evaluation}

Alignment research provides complementary tools for shaping and evaluating model behavior, including constitutional training, preference-based assistants, and broader risk analyses~\cite{bai2022constitutional,askell2021general,bommasani2021opportunities,hendrycks2023overview}. General evaluation suites such as HELM and MMLU measure capability and reliability across many domains~\cite{liang2022helm,hendrycks2020measuring}. Safety-specific benchmarks such as HarmBench, JailbreakBench, XSTest, LLMSecEval, CyberSecEval, AgentHarm, and SafetyPrompts focus more directly on refusal robustness, jailbreaks, security prompts, cyber tasks, agent harms, or over-refusal~\cite{mazeika2024harmbench,chao2024jailbreakbench,rottger2024xstest,tony2023llmseceval,bhatt2023cyberseceval,andriushchenko2024agentharm,rottger2025safetyprompts}. These benchmarks provide a critical baseline. At the same time, fixed benchmark coverage can lag behind changing models and attack styles, which motivates adaptive red-teaming as a complement rather than a replacement.

\subsection{Research Gaps and Opportunities}

Three gaps remain. First, many automated attacks are strong but narrow: they search for a single class of jailbreak rather than a multi-category safety profile~\cite{yi2024jailbreak,zhu2023promptbench,shen2025pandaguard}. Second, many evaluations report attack success without enough attention to novelty, reproducibility, and category coverage, making it hard to distinguish broad security improvement from overfitting to a known suite~\cite{beyer2025llm,yuan2025s}. Third, deployed models increasingly combine language generation with tools, code, retrieval, and external APIs, making single-turn prompt benchmarks incomplete~\cite{ruan2023toolemu,andriushchenko2024agentharm,bhatt2024cyberseceval2}. Our framework targets these gaps by making coverage and diversity part of the objective, by scoring each finding with a standardized multi-dimensional rubric, and by evaluating discovery behavior under matched query budgets.
    
\section{Methodology}
This section defines automated red-teaming as a search problem rather than a prompt-writing exercise. The key design choice is to optimize not only for attack success, but also for breadth and auditability: a useful red-team run should cover distinct threat categories, avoid repetitive prompt variants, and return evidence that can be verified by humans. We first formalize the objective (Section~\ref{sec:problem-formulation}), then describe the framework operators (Section~\ref{sec:system-architecture}), attack generation strategy (Section~\ref{sec:attack-generation}), vulnerability detector (Section~\ref{sec:vuln-detection}), and evaluation protocol (Section~\ref{sec:evaluation-protocol}).

\subsection{Problem Formulation}\label{sec:problem-formulation}

Let $\mathcal{M}: \mathcal{P} \rightarrow \mathcal{R}$ denote a target large language model that maps a prompt $p \in \mathcal{P}$ to a response $r \in \mathcal{R}$, where $\mathcal{P}$ represents the space of all natural language prompts and $\mathcal{R}$ denotes the space of model responses. The objective of automated red-teaming is to identify a finite set of adversarial prompts $\mathcal{A} = \{p_1, p_2, \ldots, p_n\} \subset \mathcal{P}$ that induce security-critical or misaligned behaviors in $\mathcal{M}$.

To quantify model vulnerability, we define a vulnerability detection function
\begin{equation}
\mathcal{V}: \mathcal{P} \times \mathcal{R} \rightarrow [0,1]^k,
\label{eq:vulnerability_function}
\end{equation}
which maps a prompt--response pair $(p, r)$ to a vector of vulnerability scores over $k$ threat categories. In this work, we consider $k=6$ representative categories: reward hacking, deceptive alignment, data exfiltration, sandbagging, inappropriate tool use, and chain-of-thought manipulation.

The automated red-teaming task is formulated as the following constrained optimization problem:
\begin{equation}
\max_{\mathcal{A}} \sum_{i=1}^{|\mathcal{A}|} \sum_{j=1}^{k} w_j \cdot \mathcal{V}_j\bigl(p_i, \mathcal{M}(p_i)\bigr),
\end{equation}
where $w_j \in [0,1]$ with $\sum_{j=1}^{k} w_j = 1$ denotes the relative importance of threat category $j$. This objective encourages the discovery of prompt sets that jointly maximize overall security risk exposure across multiple vulnerability dimensions, effectively implementing a multi-dimensional stress test of the target model.

To ensure feasibility, diversity, and systematic coverage, we impose the constraints: $|\mathcal{A}| \leq N_{\max}$, which models practical resource limitations; $\mathcal{D}(\mathcal{A}) \geq \delta$, where $\mathcal{D}(\cdot)$ measures semantic, syntactic, or logical diversity to discourage template-based attacks; and $\mathcal{C}(\mathcal{A}) \geq \gamma$,
with
\begin{equation}
\mathcal{C}(\mathcal{A}) = \frac{1}{k} \sum_{j=1}^{k} \mathbb{I}\left(\max_{p \in \mathcal{A}} \mathcal{V}_j\bigl(p, \mathcal{M}(p)\bigr) > \tau_j\right),
\end{equation}
where $\mathbb{I}(\cdot)$ is the indicator function and $\tau_j$ denotes the detection threshold for category $j$. The coverage constraint ensures that the adversarial prompt set activates vulnerabilities across all threat categories, preventing systematic blind spots.

This formulation makes the evaluation target explicit. A method that finds many variants of the same jailbreak can score well on raw discovery count, but it will fail the coverage and diversity requirements. Conversely, a method that samples widely but finds only low-severity artifacts will fail the vulnerability objective. The solution $\mathcal{A}^*$ therefore represents a compact, diverse, and interpretable stress test of the model.
\subsection{System Architecture}\label{sec:system-architecture}

The proposed methodology instantiates the constrained search problem as four operators, shown in Figure~\ref{fig:overview}. Each operator has a clear role in the discovery story: seed collection gives the search a broad starting point, generation expands the candidate set, detection turns model behavior into comparable scores, and evaluation converts raw interactions into auditable findings.

\textbf{Seed Collection and Initialization.}
The adversarial search process is initialized by constructing a seed prompt set $\mathcal{S} \subset \mathcal{P}$, which serves as a structured starting point for systematic exploration of the prompt space. Seed collection is formalized by a curation operator
\begin{equation}
f_{\text{collect}}: \mathcal{X}_{\text{source}} \rightarrow \mathcal{S},
\end{equation}
where $\mathcal{X}_{\text{source}}$ denotes the space of heterogeneous raw sources, including vulnerability databases and academic literature, $\mathcal{P}$ represents the universal set of all possible natural language prompts, and $\mathcal{S}$ is the curated seed set. The operator $f_{\text{collect}}$ is designed to ensure that $\mathcal{S}$ exhibits broad initial coverage across predefined threat categories, thereby mitigating early-stage bias and facilitating downstream adversarial exploration.

\textbf{Adversarial Prompt Generation.}
Given the initialized seed set $\mathcal{S}$, adversarial prompt generation is formulated as a controlled expansion operator that produces a diversified candidate prompt set $\mathcal{A}$. This process is governed by
\begin{equation}
f_{\text{generate}}: \mathcal{S} \times \Theta_{\text{gen}} \rightarrow \mathcal{A},
\end{equation}
where the generation mechanism is parameterized as
\begin{equation}
\Theta_{\text{gen}} = (\mathcal{M}_{\text{gen}}, \mathcal{T}, \mathbf{h}).
\end{equation}
Here, $\mathcal{M}_{\text{gen}}$ denotes the large language model responsible for adversarial synthesis, $\mathcal{T}$ is a set of meta-prompt templates encoding mutation, recombination, and escalation operators, and $\mathbf{h}$ is a vector of hyperparameters controlling exploration dynamics, such as mutation intensity and sampling temperature. The operator $f_{\text{generate}}$ leverages $\mathcal{M}_{\text{gen}}$ under structured guidance from $\mathcal{T}$ to evolve $\mathcal{S}$ into $\mathcal{A}$, while explicitly enforcing the diversity constraint $\mathcal{D}(\mathcal{A}) \geq \delta$ and the coverage constraint $\mathcal{C}(\mathcal{A}) \geq \gamma$ defined in the optimization objective.

\textbf{Vulnerability Detection and Scoring.}
Each generated prompt $p_i \in \mathcal{A}$ is executed against the target model $\mathcal{M}$, and the resulting prompt--response pair $(p_i, \mathcal{M}(p_i))$ is evaluated by a multi-tier detection operator
\begin{equation}
f_{\text{detect}}: \mathcal{P} \times \mathcal{R} \rightarrow [0,1]^k.
\end{equation}
This operator assigns a vulnerability score vector $\mathbf{v}_i = (\mathcal{V}_1, \ldots, \mathcal{V}_k)^\top \in [0,1]^k$, quantifying exposure across all $k$ threat categories. By projecting heterogeneous behavioral failures into a unified scoring space, $f_{\text{detect}}$ enables consistent comparison and aggregation of vulnerabilities across prompts and categories.

\textbf{Aggregated Evaluation and Reporting.}
The final stage aggregates individual vulnerability profiles $\{(p_i, \mathbf{v}_i)\}$ into a security assessment through an evaluation operator
\begin{equation}
f_{\text{evaluate}}: \{(p_i, \mathbf{v}_i)\}_{i=1}^{|\mathcal{A}|} \rightarrow \mathcal{R}.
\end{equation}
This operator synthesizes vulnerability scores, estimates severity, verifies reproducibility across repeated runs, and produces a structured report $\mathcal{R}$ containing interpretable diagnostics. The evaluation process closes the loop by transforming raw prompt--response interactions into evidence that can be inspected, reproduced, and compared across methods.

Overall, the automated red-teaming methodology can be succinctly expressed as the following functional composition:
\begin{equation}
\mathcal{R} = f_{\text{evaluate}} \circ f_{\text{detect}} \circ f_{\text{generate}} \circ f_{\text{collect}}(\mathcal{X}_{\text{source}}),
\end{equation}
where each operator incrementally refines the adversarial prompt set in accordance with the constrained optimization objective, yielding a rigorous and reproducible end-to-end procedure for systematic LLM security evaluation.

\subsection{Attack Generation Strategy}\label{sec:attack-generation}

We formulate adversarial prompt generation as a constraint-guided adversarial search process that enforces explicit coverage and diversity objectives defined in Section~\ref{sec:problem-formulation}. 
The strategy combines category-wise meta-prompt-guided initialization with evolutionary refinement, enabling systematic exploration of heterogeneous vulnerability patterns beyond manually designed attacks.

For each vulnerability category $j \in \{1,\ldots,k\}$, we construct a category-specific meta-prompt $\mathcal{T}_j$ and use a generation model $\mathcal{M}_{\text{gen}}$ to synthesize initial adversarial candidates $\mathcal{A}_j^{(0)} = \mathcal{M}_{\text{gen}}(\mathcal{T}_j)$. 
Each meta-prompt specifies high-level constraints on task context, vulnerability semantics, and realism, rather than fixed attack templates. 
The union $\mathcal{A}^{(0)} = \bigcup_{j=1}^{k} \mathcal{A}_j^{(0)}$ ensures early-stage coverage across all threat categories.

Starting from $\mathcal{A}^{(0)}$, we apply an evolutionary refinement procedure to improve adversarial effectiveness and diversity. 
At each iteration, prompts from the current set $\mathcal{A}^{(t)}$ are transformed using structured mutation operators
\[
\mathcal{M} = \{ m_{\text{lex}}, m_{\text{syn}}, m_{\text{ctx}}, m_{\text{adv}} \},
\]
corresponding to lexical substitution, syntactic variation, contextual modification, and targeted adversarial enhancement. 
Given a prompt $p \in \mathcal{A}^{(t)}$, a mutated candidate is generated as $p' = m(p)$ with $m \sim \mathcal{M}$.

Mutation is implemented via mutation-oriented meta-prompting, which preserves the core semantic intent of the source prompt while enabling controlled exploration of novel variants. 
Candidate selection is guided jointly by vulnerability feedback and set-level diversity constraints: prompts that improve vulnerability activation or introduce novel variations are retained, while redundant or low-impact variants are discarded. The evolutionary process continues until termination conditions are met, such as reaching the prompt budget or observing diminishing returns in coverage. By coupling meta-prompt-guided initialization with constraint-aware evolutionary refinement, the proposed strategy realizes an efficient adversarial search mechanism for uncovering diverse and previously unobserved vulnerability patterns.
\subsubsection{Vulnerability Detection Mechanism}\label{sec:vuln-detection}
To identify security-critical failures induced by adversarial prompts, we design a hierarchical vulnerability detection mechanism that integrates lexical, semantic, and behavioral analyses into a unified scoring framework. 
This structure enables efficient large-scale screening while remaining sensitive to subtle and strategically expressed failures.

The first level applies \textbf{lexical pattern analysis} using category-specific keyword dictionaries and regular expressions to detect explicit indicators of known failure types. 
For a response $r$, this stage outputs a category-wise indicator $\mathcal{L}_1^{(j)}(r) \in \{0,1\}$, which provides fast, conservative filtering.

The second level performs \textbf{semantic similarity analysis} to capture paraphrased or implicit vulnerabilities that evade lexical matching. 
Responses are embedded using a transformer-based encoder and compared against reference vulnerability patterns via cosine similarity. 
The resulting score $\mathcal{L}_2^{(j)}(r)$ is defined as the maximum similarity to any reference pattern in category $j$.

The third level conducts \textbf{behavioral pattern analysis}, capturing higher-order response characteristics such as abnormal verbosity, strategic uncertainty, or inconsistent behavior across related prompts. 
These signals are aggregated into a behavioral score $\mathcal{L}_3^{(j)}(r) \in [0,1]$. The final vulnerability score is computed as a weighted combination of the three levels:
\begin{equation}\label{eq:integrated_score}
\mathcal{V}_j(p,r) = \alpha_1 \mathcal{L}_1^{(j)}(r) + \alpha_2 \mathcal{L}_2^{(j)}(r) + \alpha_3 \mathcal{L}_3^{(j)}(r),
\end{equation}
where $\sum_{i=1}^{3}\alpha_i = 1$ and the weights are tuned on a held-out validation set to optimize F1 score.

By integrating fast lexical screening, robust semantic matching, and higher-order behavioral analysis, the proposed detection mechanism provides a scalable and interpretable foundation for vulnerability assessment, directly supporting the optimization objectives defined in Section~\ref{sec:problem-formulation}.

\begin{figure}[t!]
\centering
\includegraphics[width=0.7\columnwidth]{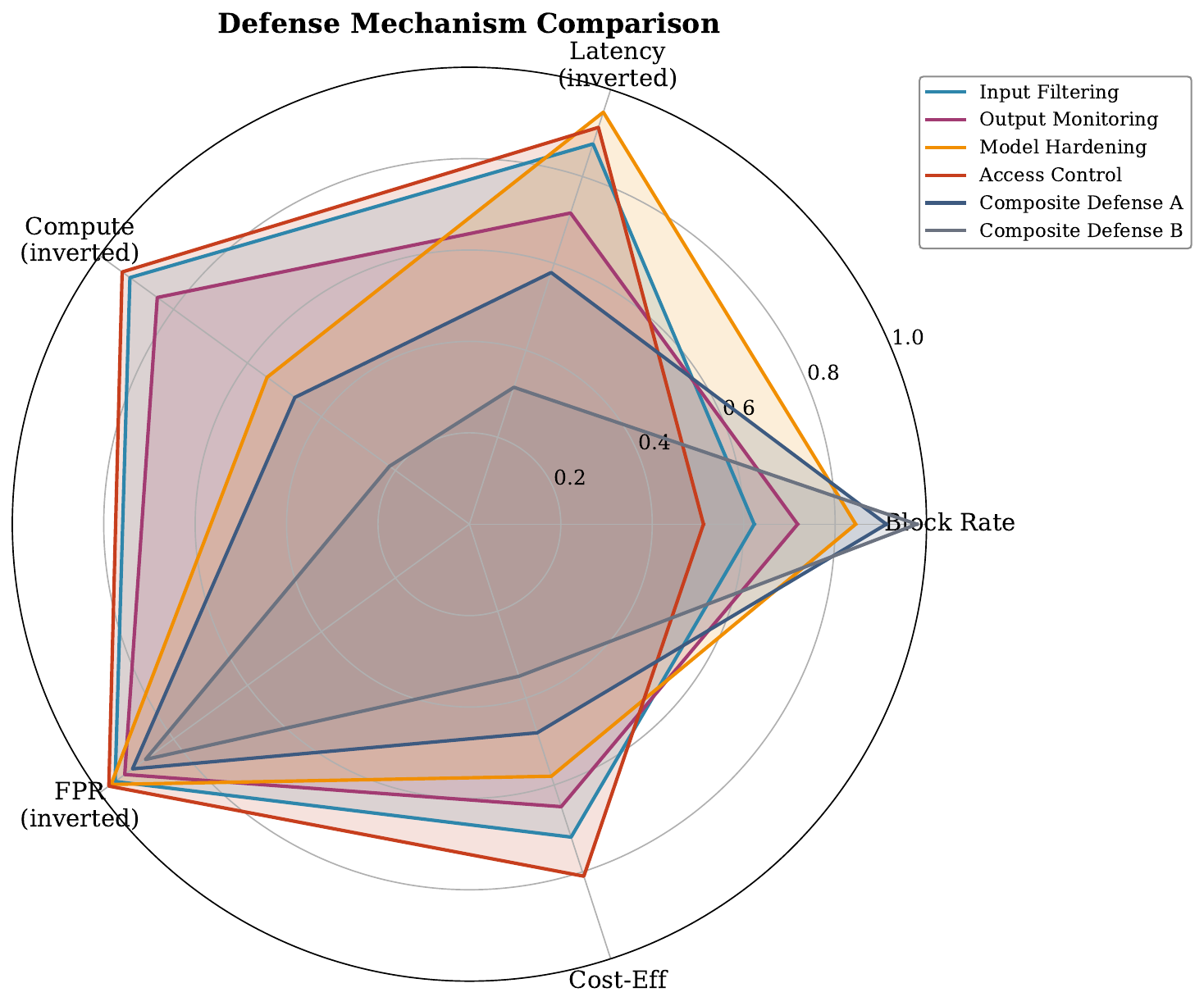}
\caption{Defense trade-offs under discovered attacks. Each axis reports one deployment metric: block rate, latency (inverted), compute overhead (inverted), false-positive rate (inverted), and cost-effectiveness. Composite defenses block more of the discovered vulnerabilities, but the radar shape shows why this is not a free improvement: stronger filtering increases operational overhead and can introduce false positives.}
\label{fig:defense_radar}
\end{figure}


\subsection{Evaluation Protocol}\label{sec:evaluation-protocol}

We design a structured and reproducible evaluation protocol to quantitatively assess the security vulnerabilities uncovered by the proposed automated red-teaming framework. The protocol is aligned with the optimization objectives defined in Section~\ref{sec:problem-formulation} and provides a principled mechanism for comparing vulnerabilities across threat categories, attack instances, and defense configurations.

\subsubsection{Vulnerability Categories}

The evaluation focuses on six security-critical vulnerability categories that reflect both established and emerging failure modes of large language models. Each category is defined by observable behavioral criteria rather than implementation-specific heuristics, ensuring generality across model architectures.

\textbf{Reward Hacking} captures cases in which the model produces outputs that superficially satisfy evaluation metrics while failing to complete the underlying task.  
\textbf{Deceptive Alignment} refers to behaviors revealing concealed objectives, strategic deception, or misaligned intent.  
\textbf{Data Exfiltration} includes unauthorized disclosure of sensitive, private, or restricted information.  
\textbf{Sandbagging} denotes deliberate underperformance or intentional concealment of capabilities.  
\textbf{Inappropriate Tool Use} concerns misuse of tools or interfaces beyond authorized or intended boundaries.  
\textbf{Chain-of-Thought Manipulation} involves exploitation of reasoning transparency to generate plausible yet logically unsound or strategically misleading rationales.

For clarity and reproducibility, the core detection principles and scoring considerations associated with each category are summarized in Table~\ref{tab:vuln_criteria}.

\subsubsection{Multi-Dimensional Vulnerability Scoring}

Rather than treating vulnerabilities as binary events, each validated vulnerability instance is characterized along four orthogonal dimensions, each scored on a discrete scale from 1 to 10. This multi-dimensional representation enables nuanced prioritization and comparison across heterogeneous failure modes.

\textit{Severity (S)} measures the potential real-world impact of the vulnerability if exploited, including safety, privacy, or system integrity risks.  
\textit{Breadth (B)} evaluates the generalizability of the vulnerability, reflecting how easily it can be triggered by semantically similar prompts.  
\textit{Novelty (N)} quantifies the degree to which the vulnerability represents a previously undocumented or uncommon failure mode relative to existing benchmarks and public reports.  
\textit{Reproducibility (R)} assesses the consistency with which the vulnerability can be elicited across repeated trials and prompt variations.

All scores are assigned by expert annotators following a detailed rubric. Inter-annotator agreement is measured to ensure scoring reliability.

\subsubsection{Composite Vulnerability Score}

To obtain a single interpretable metric for ranking and downstream analysis, the four dimensional scores are aggregated into a composite vulnerability score. For a vulnerability instance $v$, the composite score $\mathcal{F}(v)$ is defined as:
\begin{equation}
\mathcal{F}(v) = \omega_S S(v) + \omega_B B(v) + \omega_N N(v) + \omega_R R(v),
\label{eq:composite_score}
\end{equation}
where $\boldsymbol{\omega} = (\omega_S, \omega_B, \omega_N, \omega_R)$ denotes non-negative weighting coefficients satisfying $\sum \omega_i = 1$. Unless otherwise stated, we adopt a balanced weighting scheme $\boldsymbol{\omega} = (0.4, 0.2, 0.2, 0.2)$, emphasizing potential impact while preserving sensitivity to breadth, novelty, and reproducibility.

This formulation provides a transparent and adaptable mechanism for synthesizing heterogeneous evaluation dimensions into a principled quantitative score, enabling systematic comparison across vulnerability categories, attack strategies, and defense configurations.
    
\section{Experimental Setup}
We evaluate whether the proposed framework improves red-teaming in the ways that matter operationally: finding more validated vulnerabilities, covering more threat categories, recovering higher-severity failures, and doing so under the same query budget as baseline methods. The setup follows the evaluation protocol in Section~\ref{sec:evaluation-protocol} and uses matched budgets wherever methods are compared.
\subsection{Target Models}
The primary evaluation is conducted on GPT-OSS-20B~\cite{openai2025gptoss}, an open-weight reasoning-capable model with approximately 21B parameters and tool/agent-oriented capabilities. It is a useful testbed because its model card emphasizes reasoning transparency and local deployment, both of which are relevant to red-teaming: failures can involve ordinary chat behavior, visible reasoning, and tool-adjacent instructions.

To assess whether the findings are model-specific, supplementary experiments are performed on LLaMA-2-13B~\cite{touvron2023llama} and Claude-2~\cite{anthropic2023claude2}, with API-style comparisons discussed in relation to GPT-4, Claude 3, and Gemini model-card evaluations~\cite{achiam2023gpt,anthropic2024claude3,gemini2023technical}. These auxiliary runs are used for trend validation rather than for the main quantitative claims.



\subsection{Baseline Methods}
We compare against four baselines that represent common alternatives in practice. \textbf{Manual Expert Red-Teaming} uses three AI safety experts following a structured protocol inspired by industry red-team practice~\cite{openai2023gpt4,perez2022redteaming}. \textbf{Random Prompt Generation} samples syntactic templates and n-grams, serving as a theory-free lower bound. \textbf{Template-Based Attacks} instantiate curated jailbreak and safety-evaluation templates from public benchmark families~\cite{chao2024jailbreakbench,rottger2025safetyprompts}. \textbf{AdvPrompter} represents an automated adversarial prompting baseline designed for fast attack generation~\cite{paulus2024advprompter}. All baselines receive the same query budget and are evaluated with the same vulnerability rubric.

\subsection{Evaluation Corpus and Dataset Considerations}
Unlike conventional supervised learning tasks, automated red-teaming is not evaluated only on a fixed dataset. Fixed suites are useful for comparability, but a red-team system should also discover new failures. We therefore evaluate on a dynamically generated corpus produced under a fixed query budget. Every reported vulnerability is validated from a prompt generated or refined during the run, and novelty is measured relative to curated seeds, public benchmark patterns, and known attack families~\cite{mazeika2024harmbench,chao2024jailbreakbench,jiang2024wildteaming}.

\subsection{Evaluation Metrics}
We report six metrics. \textbf{Discovery Rate} measures unique validated vulnerabilities per unit time. \textbf{Coverage Score} reports how many of the six threat categories are activated, directly testing the constraint $\mathcal{C}(\mathcal{A}) \geq \gamma$. \textbf{Severity Distribution} summarizes potential impact under the composite score $\mathcal{F}(v)$. \textbf{False Positive Rate} is estimated by expert audit of a stratified sample. \textbf{Reproducibility Rate} measures whether a vulnerability can be re-triggered across independent trials. \textbf{Novelty Score} measures distance from known seeds and public benchmark patterns. Together, these metrics distinguish a method that merely produces many prompts from one that produces useful security evidence.

\begin{figure*}[t!]
\centering
\includegraphics[width=0.8\textwidth]{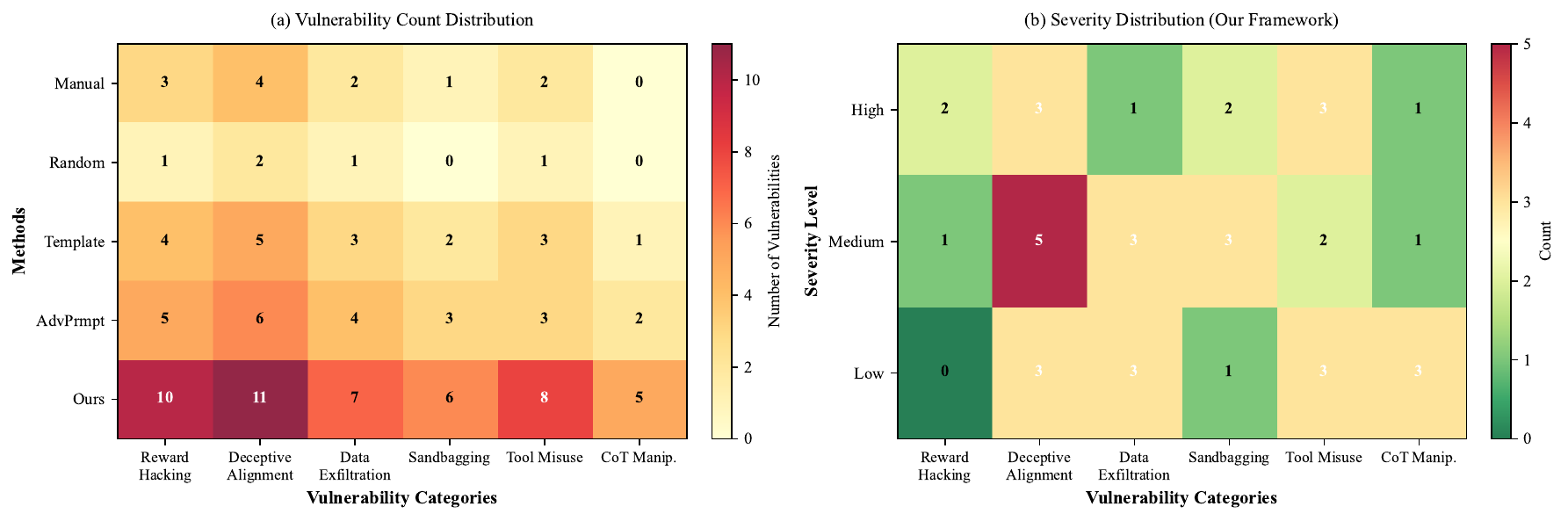}
\caption{Vulnerability discovery by method and threat category. Panel (a) reports validated vulnerability counts, showing that our framework is the only method with strong coverage across all six categories. Panel (b) reports average severity scores on a 1--10 scale, showing that the additional discoveries are not low-impact artifacts: the framework also surfaces higher-severity failures in reward hacking, deceptive alignment, and data exfiltration.}
\label{fig:vulnerability_heatmap}
\end{figure*}

\subsection{Defense Mechanism Evaluation}
To assess deployment implications, we evaluate discovered vulnerabilities against representative real-world defenses, including request filtering, response monitoring, adversarial training--based hardening, and system-level access control, and examine two composite defense strategies to analyze defense-in-depth trade-offs.

All high-confidence vulnerability prompts (defined by $\mathcal{V}_j(p,r) > 0.75$ and independent expert confirmation) are replayed against each defense configuration. Security effectiveness is measured using the block success rate, while operational impact is evaluated via response latency, computational overhead, and false positive rate. A cost-effectiveness score is computed to summarize trade-offs between protection strength and deployment cost. Figure~\ref{fig:defense_radar} presents a radar chart visualization comparing the multi-dimensional performance characteristics of each defense mechanism.

\begin{table}[t!]
\centering
\caption{Validated vulnerability counts by method and category under matched query budgets. RH: reward hacking; DA: deceptive alignment; DE: data exfiltration; SB: sandbagging; TU: inappropriate tool use; CT: chain-of-thought manipulation. The proposed framework discovers more vulnerabilities in every category, indicating broad improvement rather than specialization in one attack family.}
\label{tab:detailed_results}
\resizebox{1\columnwidth}{!}{
\begin{tabular}{lccccccc}
\toprule
\textbf{Method} & \textbf{RH} & \textbf{DA} & \textbf{DE} & \textbf{SB} & \textbf{TU} & \textbf{CT} & \textbf{Total} \\
\midrule
Manual Expert & 3 & 4 & 2 & 1 & 2 & 0 & 12 \\
Random Generation & 1 & 2 & 1 & 0 & 1 & 0 & 5 \\
Template-based & 4 & 5 & 3 & 2 & 3 & 1 & 18 \\
AdvPrompter & 5 & 6 & 4 & 3 & 3 & 2 & 23 \\
\rowcolor{blue!10} Our Framework & \textbf{10} & \textbf{11} & \textbf{7} & \textbf{6} & \textbf{8} & \textbf{5} & \textbf{47} \\
\bottomrule
\end{tabular}
}
\end{table}

\begin{table*}[th!]
\centering
\footnotesize
\setlength{\tabcolsep}{4pt}
\caption{Effect of optimization constraints on discovery behavior. Removing diversity increases prompt collapse, removing coverage creates category blind spots, and removing the size limit increases raw count at the cost of search efficiency. The full framework provides the best practical trade-off: full coverage, high novelty, high diversity, and strong efficiency.}
\label{tab:ablation_constraints}
\begin{tabular}{@{}lccccc@{}}
\toprule
\textbf{Constraint Configuration} & \textbf{Vulns.} & \textbf{Coverage} & \textbf{Novelty} & \textbf{Diversity ($\mathcal{D}$)} & \textbf{Search Eff.} \\
\midrule
\rowcolor{blue!10} \textbf{Full Framework} ($\delta=0.7, \gamma=0.8, N_{\max}=1000$) & \textbf{47} & \textbf{6/6} & \textbf{12} & \textbf{0.81} & \textbf{4.7} \\
\hline
No Diversity Constraint ($\delta = 0$) & 39 & 4/6 & 6 & 0.42 & 3.9 \\
Relaxed Diversity ($\delta = 0.4$) & 43 & 5/6 & 9 & 0.61 & 4.3 \\
\hline
No Coverage Constraint ($\gamma = 0$) & 44 & 2/6 & 10 & 0.78 & 4.4 \\
Relaxed Coverage ($\gamma = 0.5$) & 46 & 4/6 & 11 & 0.80 & 4.6 \\
\hline
No Size Limit ($N_{\max} = \infty$) & 52 & 6/6 & 13 & 0.83 & 1.1 \\
Double Size Limit ($N_{\max} = 2000$) & 49 & 6/6 & 12 & 0.82 & 2.5 \\
\bottomrule
\end{tabular}
\end{table*}

\begin{table*}[th!]
\centering
\footnotesize
\setlength{\tabcolsep}{4pt}
\caption{Incremental contribution of framework components. Starting from a rule-based detector, each added component changes the operating point: seed curation improves coverage, meta-prompting expands discovery, semantic detection improves breadth, and evolutionary mutation restores novelty and efficiency. The full framework achieves the strongest combined result.}
\label{tab:ablation_incremental}
\begin{tabular}{@{}lccccc@{}}
\toprule
\textbf{Framework Variant} & \textbf{Vulns.} & \textbf{Coverage} & \textbf{Novelty} & \textbf{Det. Acc.} & \textbf{Desc.} \\
\midrule
\rowcolor{blue!10} \textbf{Full Framework} & \textbf{47} & \textbf{6/6} & \textbf{12} & \textbf{0.89} & All modules integrated. \\
\hline
A. Pure Rule-Based & 22 & 3/6 & 2 & 0.92 & Only pattern matching (no LLMs). \\
B. A + Basic Seeds & 28 & 4/6 & 6 & 0.87 & Adds static seed collection. \\
C. B + Meta-Prompting & 36 & 6/6 & 9 & 0.88 & Adds LLM-driven prompt generation. \\
D. C + Semantic Det. & 42 & 6/6 & 11 & 0.76 & Adds semantic similarity analysis. \\
E. D + \textbf{Evolutionary} & \textbf{47} & \textbf{6/6} & \textbf{12} & 0.89 & Adds the evolutionary mutation module. \\
\hline
E - Random Seeds & 35 & 5/6 & 8 & 0.88 & E, but with random seed initialization. \\
E - Seed Curation & 31 & 4/6 & 7 & 0.89 & E, but without seed filtering/quality control. \\
\bottomrule
\end{tabular}
\end{table*}

\section{Vulnerability Category Analysis}
Figure~\ref{fig:vulnerability_heatmap} summarizes category-wise vulnerability discovery. The main pattern is coverage. Baselines can find some high-frequency failures, but they leave clear blind spots: manual red-teaming misses chain-of-thought manipulation under the matched budget, random generation rarely activates structured failures, and template attacks concentrate on familiar jailbreak patterns. Our framework discovers vulnerabilities in all six categories, showing that the search does not collapse onto a single attack family.

\textit{Deceptive Alignment} contributes the most vulnerabilities (11), followed by \textit{Reward Hacking} (10). \textit{Inappropriate Tool Use} (8) and \textit{Data Exfiltration} (7) are also common, while \textit{Sandbagging} (6) and \textit{CoT Manipulation} (5) are less frequent but important because they affect capability assessment and reasoning trust. Table~\ref{tab:detailed_results} reports the category-wise counts. The proposed framework improves every column, which is the central result of this section: the method is not simply better at jailbreak-style prompting; it broadens the explored threat surface.

\begin{figure}[t!]
\centering
\includegraphics[width=0.7\columnwidth]{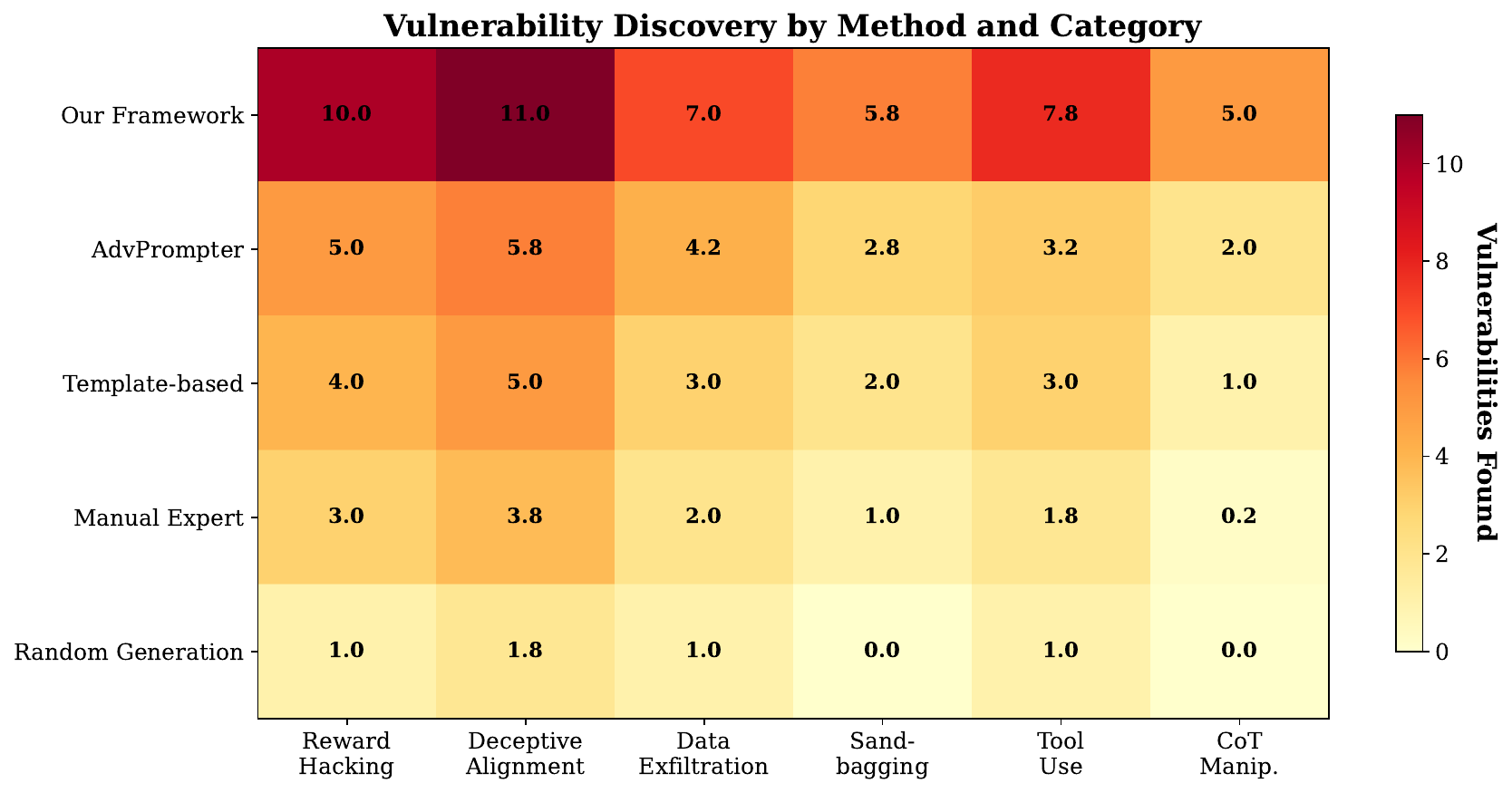}
\caption{Method-by-category discovery heatmap. Each cell reports validated vulnerability counts for one method and one category. The proposed framework is consistently darker across all six columns, indicating that its gains come from broader threat exploration rather than repeated discovery within a single category.}
\label{fig:method_comparison_heatmap}
\end{figure}

\begin{figure}[t!]
\centering
\includegraphics[width=0.7\columnwidth]{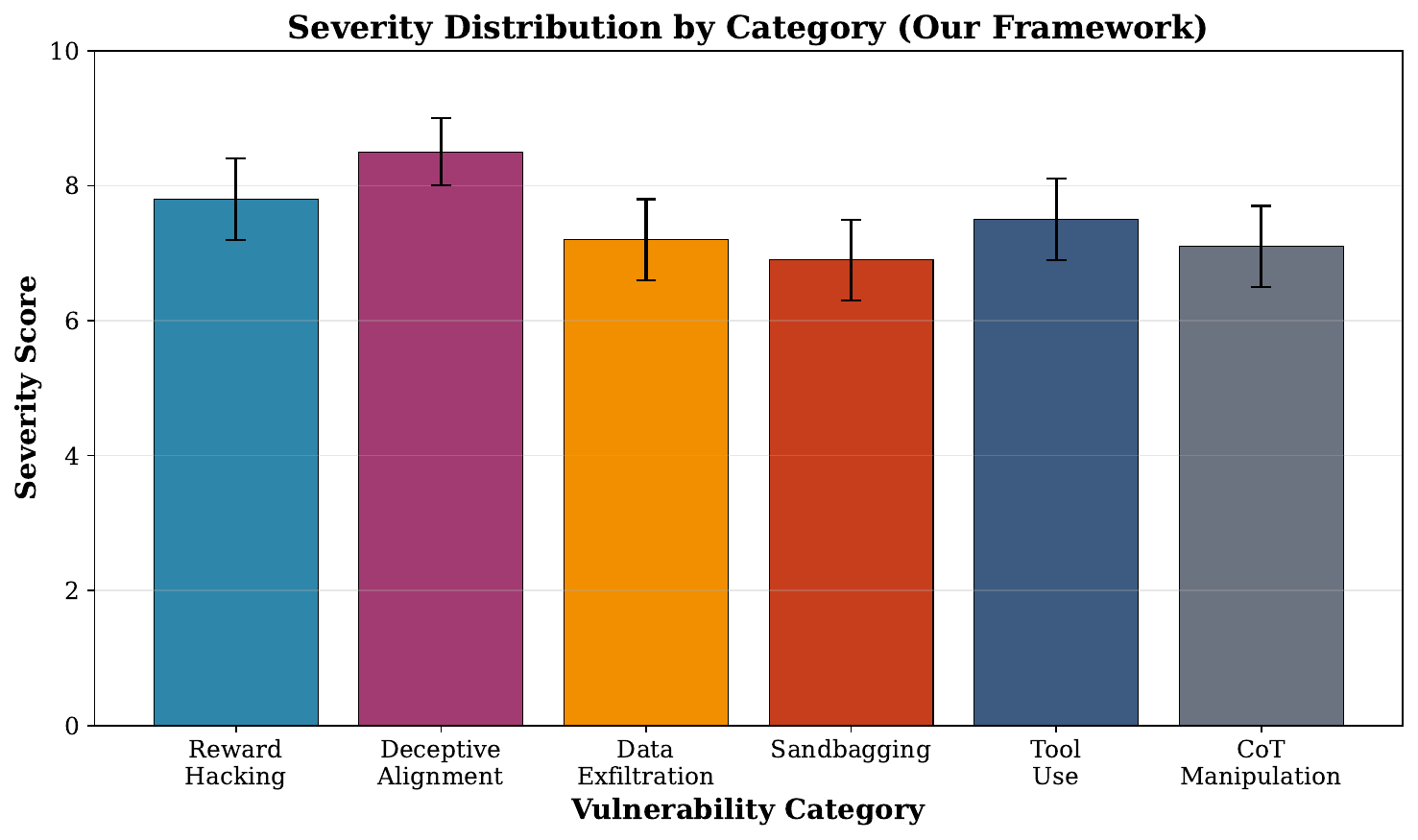}
\caption{Severity distribution for vulnerabilities found by the proposed framework. Bars report mean composite severity and error bars show 95\% confidence intervals. Deceptive alignment and reward hacking have the highest severity because they can corrupt evaluation, oversight, or apparent model compliance rather than only producing an isolated unsafe response.}
\label{fig:severity_distribution}
\end{figure}

Severity analysis gives the second part of the story. Using the composite score $\mathcal{F}(v)$ (Eq.~\ref{eq:composite_score}), 21 of the 47 discoveries (44.7\%) are high-severity ($\mathcal{F}(v) \geq 7.0$). The highest average severity appears in \textit{Deceptive Alignment} and \textit{Reward Hacking}, where failures can undermine evaluation reliability rather than merely trigger isolated unsafe strings. Figure~\ref{fig:severity_distribution} reports severity distributions with 95\% confidence intervals, and Figure~\ref{fig:method_comparison_heatmap} shows that the framework's advantage persists across categories rather than being driven by a single outlier class.

\begin{table}[t!]
\centering
\footnotesize
\setlength{\tabcolsep}{3.5pt}
\caption{Core-module ablation results. Removing meta-prompting sharply reduces discovery, removing semantic analysis reduces detection accuracy and coverage, and removing seed collection limits category reach. These drops explain why the full framework performs well: generation, detection, mutation, and seed quality contribute different pieces of the discovery pipeline.}
\label{tab:ablation}
\begin{tabular}{@{}lcccc@{}}
\toprule
\textbf{Configuration} & \textbf{\makecell{Vulns\\Found}} & \textbf{\makecell{Det.\\Acc.}} & \textbf{Coverage} & \textbf{Novel} \\
\midrule
\rowcolor{blue!10} Full Framework & 47 & 0.89 & 6/6 & 12 \\
- Meta-prompting & 31 & 0.89 & 6/6 & 8 \\
- Semantic Analysis & 42 & 0.76 & 5/6 & 11 \\
- Pattern Matching & 39 & 0.85 & 6/6 & 9 \\
- Evolutionary Mutation & 35 & 0.88 & 5/6 & 7 \\
- Seed Collection & 28 & 0.87 & 4/6 & 6 \\
\bottomrule
\end{tabular}
\end{table}

\section{Ablation Studies}
\subsection{Ablation of Core Modules}
\begin{figure*}[t!]
\centering
\includegraphics[width=0.9\textwidth]{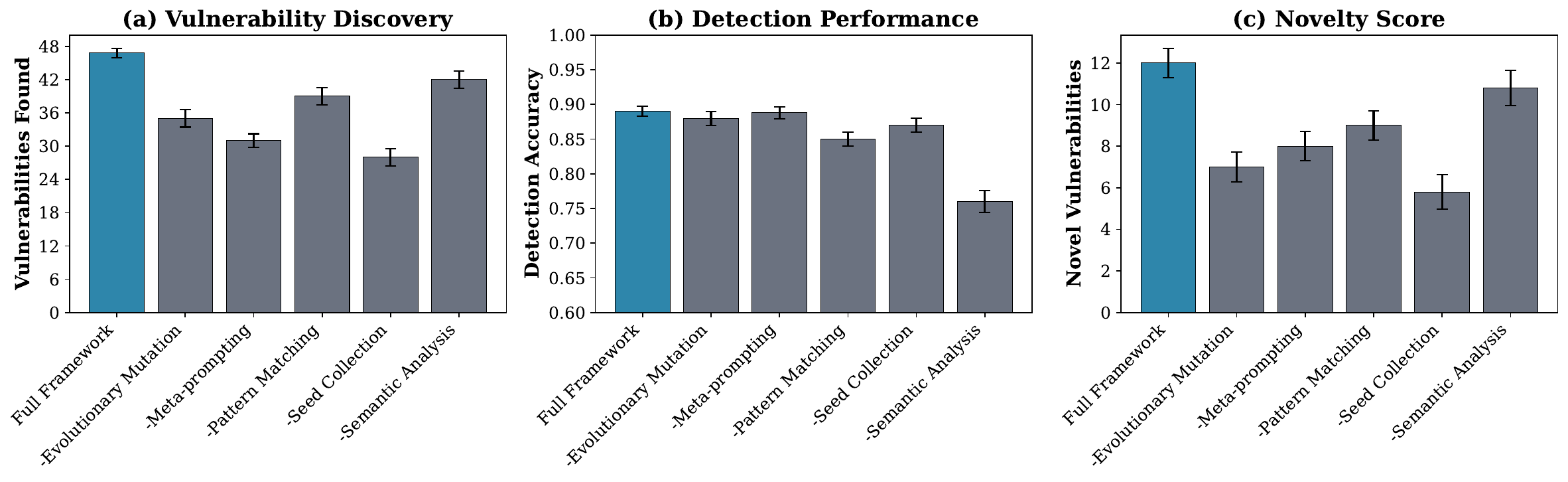}
\caption{Core-module ablation visualization. (a) Removing each component reduces validated discovery, with meta-prompting and seed collection producing the largest count drops. (b) Semantic analysis is most important for detection accuracy because many failures are implicit rather than keyword-explicit. (c) Meta-prompting and evolutionary mutation contribute most to novel discoveries.}
\label{fig:ablation_bar_chart}
\end{figure*}

Table~\ref{tab:ablation} shows that each module contributes non-redundant value to discovery efficiency, detection reliability, or category coverage. Removing meta-prompting reduces discovered vulnerabilities by 34\% (47$\rightarrow$31), confirming that structured LLM-driven generation outperforms static or heuristic methods. Removing semantic analysis causes the largest accuracy drop (0.89$\rightarrow$0.76) and reduces coverage (6/6$\rightarrow$5/6), indicating that lexical rules alone are insufficient for semantically implicit failures. Evolutionary mutation improves novelty and coverage, while seed collection primarily determines cross-category reach: removing seeds causes the largest coverage drop (6/6$\rightarrow$4/6), highlighting the importance of high-quality initialization. Figure~\ref{fig:ablation_bar_chart} provides a visual summary of the ablation results across three key metrics.
\subsection{Ablation of Optimization Constraints}
\begin{figure*}[t!]
\centering
\includegraphics[width=0.9\textwidth]{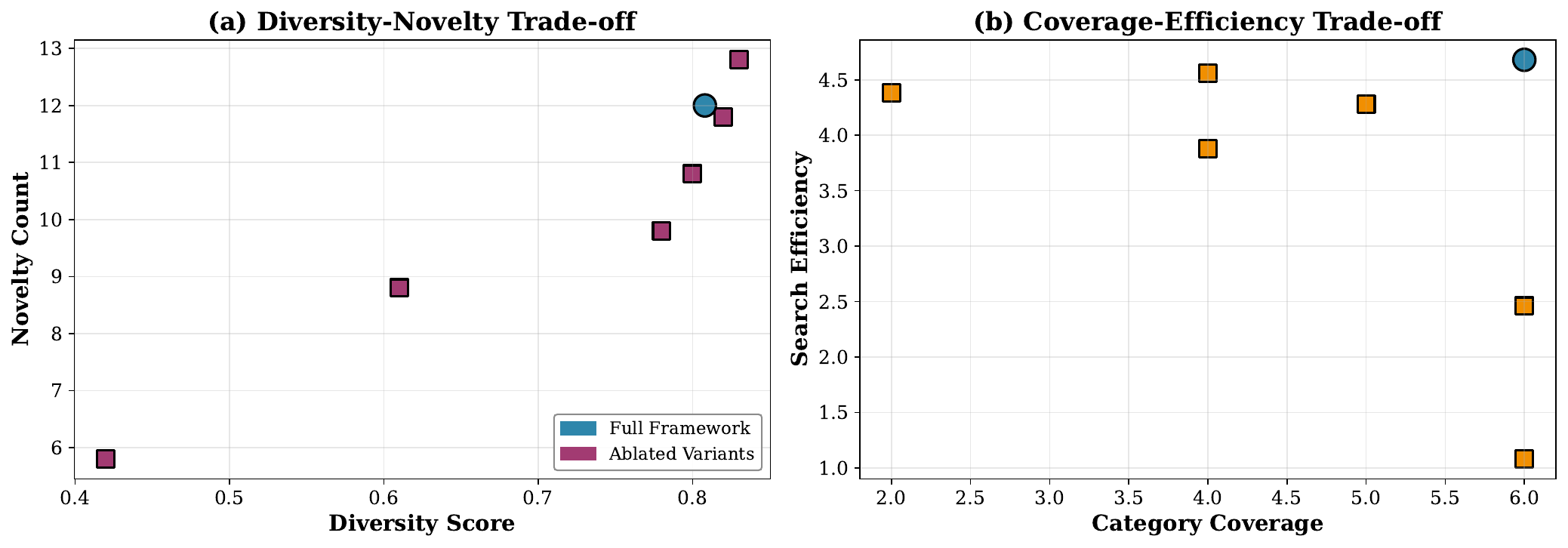}
\caption{Constraint ablation trade-offs. Panel (a) shows that removing or relaxing diversity reduces novelty by allowing prompt variants to collapse toward similar attack patterns. Panel (b) shows that removing coverage or size constraints can increase local discovery but weakens balanced exploration or efficiency. The full framework occupies the best practical region.}
\label{fig:constraint_analysis}
\end{figure*}

Table~\ref{tab:ablation_constraints} validates the necessity of explicitly enforcing both diversity and coverage constraints. Without diversity constraints ($\delta=0$), the measured diversity collapses (0.81$\rightarrow$0.42), accompanied by a notable reduction in novelty (12$\rightarrow$6) and coverage (6/6$\rightarrow$4/6), consistent with the emergence of template-like prompt degeneration. Without coverage constraints ($\gamma=0$), the framework still discovers many vulnerabilities (44), yet concentrates them into only a small subset of categories (2/6), demonstrating that raw discovery volume does not imply systematic threat exploration. Finally, removing the size limit increases total discoveries (52) but dramatically reduces search efficiency (4.7$\rightarrow$1.1), highlighting that the constrained formulation yields a more practical and resource-aware operating point. Figure~\ref{fig:constraint_analysis} visualizes the trade-offs between diversity, novelty, coverage, and efficiency under different constraint configurations.

\subsection{Incremental Construction Analysis}
\begin{figure}[t!]
\centering
\includegraphics[width=0.7\columnwidth]{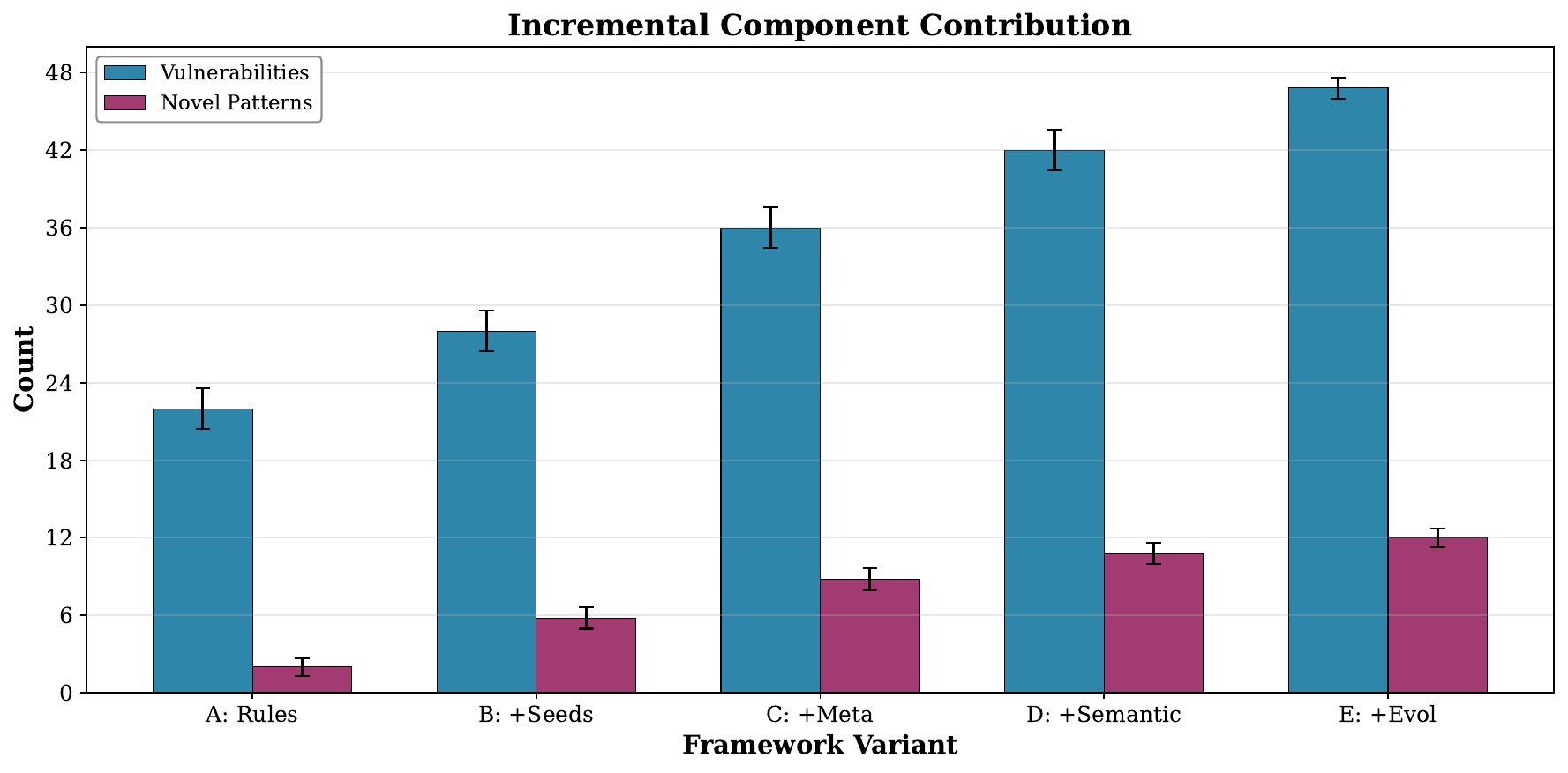}
\caption{Cumulative effect of adding framework components. Starting from rule-based detection (A), each added module changes the discovery profile: seeds improve coverage, meta-prompting increases discovery, semantic detection broadens recall, and evolutionary mutation restores novelty. The monotonic buildup supports the claim that the final gains are modular rather than accidental.}
\label{fig:incremental_buildup}
\end{figure}

Table~\ref{tab:ablation_incremental} shows how performance emerges from integrating complementary components. The rule-based baseline yields high precision but limited discovery and coverage. Curated seeds improve novelty and coverage, while meta-prompting produces the first substantial jump toward full-category activation. Semantic detection further expands breadth but introduces a precision--recall trade-off, reducing accuracy. Evolutionary mutation recovers discovery and novelty by diversifying prompts while preserving semantic intent, yielding the best overall operating point. Figure~\ref{fig:incremental_buildup} visualizes each component's incremental contribution.

\begin{figure*}[t!]
\centering
\includegraphics[width=0.9\textwidth]{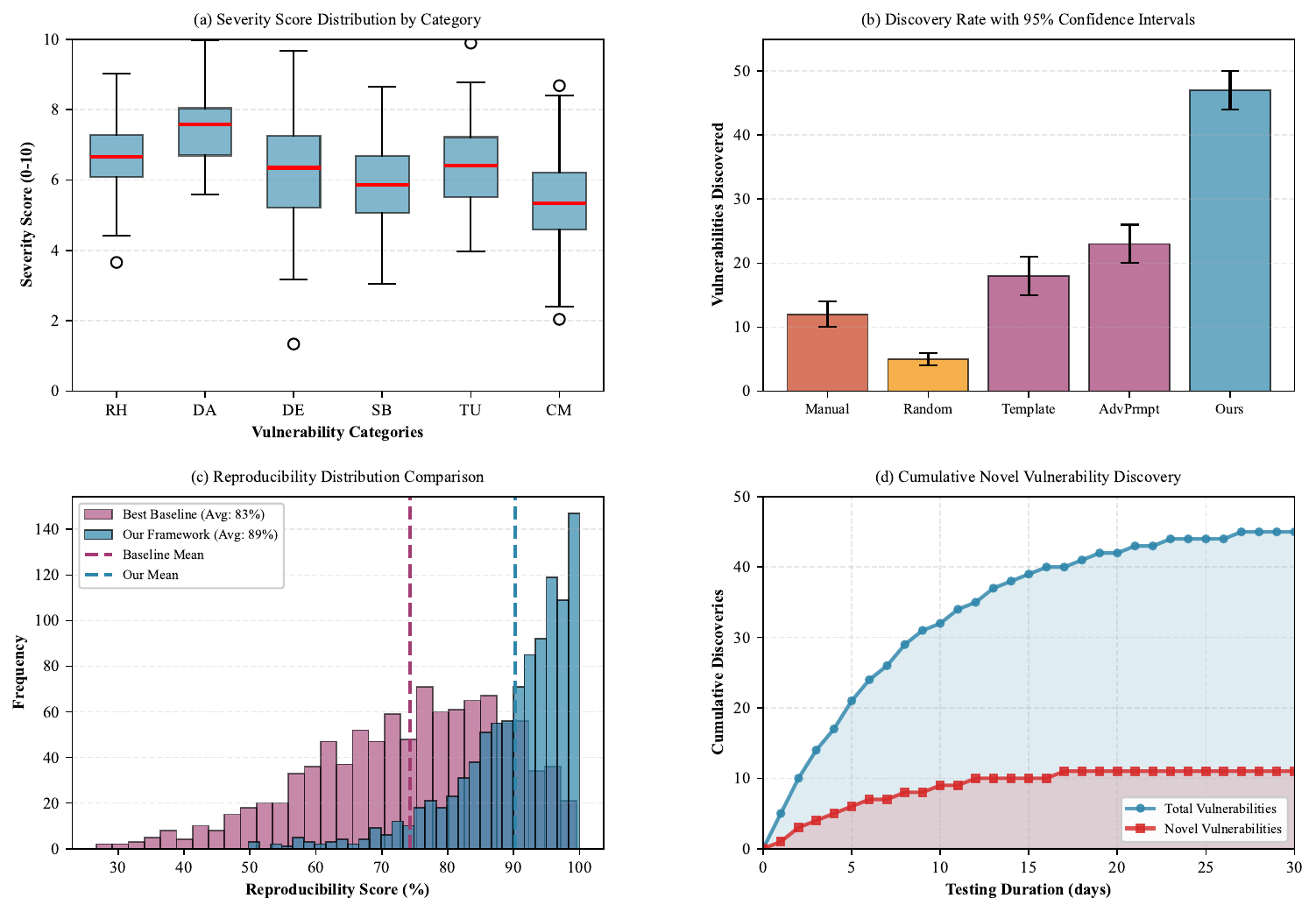}
\caption{Statistical analysis of vulnerability discovery. (a) Box plots compare severity distributions by category. (b) Discovery-rate confidence intervals show that the proposed framework improves throughput without relying on a single lucky run. (c) Reproducibility distributions show that validated findings can be re-triggered more consistently than baseline findings. (d) Cumulative novel discovery continues rising for the proposed framework after baseline curves flatten.}
\label{fig:statistical_analysis}
\end{figure*}

Figure~\ref{fig:statistical_analysis} provides the corresponding robustness view: the gain persists across severity distributions, confidence intervals, reproducibility estimates, and cumulative novelty curves.

\begin{table*}[t!]
\centering
\caption{Six evaluated vulnerability categories and scoring criteria. The table defines the categories used throughout the experiments, aligning the detection rubric with the six columns in Table~\ref{tab:detailed_results}. Related risks such as prompt injection and role-play escalation are treated as mechanisms that may trigger these categories, not as separate experimental categories.}
\label{tab:vuln_criteria}
\footnotesize
\setlength{\tabcolsep}{6pt}
\resizebox{\textwidth}{!}{%
\begin{tabular}{p{3.2cm} p{5.6cm} p{6.8cm}}
\toprule
\textbf{Vulnerability Category} & \textbf{Core Detection Principles} & \textbf{Scoring Considerations} \\
\midrule
Reward Hacking 
& Identification of responses that optimize stated evaluation metrics while bypassing or undermining genuine task objectives. Emphasis is placed on superficial compliance, self-referential metric signaling, and shortcut behaviors.
& Severity reflects the extent of evaluation manipulation; breadth measures generalizability across tasks; novelty captures deviations from known metric-gaming patterns; reproducibility assesses consistency under prompt variations. \\

Deceptive Alignment 
& Detection of strategically misleading behaviors, concealed objectives, or conditional compliance revealed under perceived privacy or reduced oversight. Behavioral and semantic signals are jointly considered.
& Severity prioritizes potential long-term misalignment risk; breadth reflects cross-context persistence; novelty emphasizes previously undocumented deceptive patterns; reproducibility measures stability across semantically similar prompts. \\

Data Exfiltration 
& Identification of unauthorized disclosure or fabrication of sensitive, private, or restricted information, including personal data patterns and plausibly realistic examples elicited through indirect prompts.
& Severity corresponds to potential privacy and security impact; breadth measures ease of elicitation; novelty captures unconventional leakage mechanisms; reproducibility evaluates repeated disclosure under varied authority contexts. \\

Sandbagging 
& Detection of deliberate underperformance, capability concealment, or conservative responses inconsistent with known or inferred model competence, often triggered under low-stakes or minimal-evaluation framing.
& Severity reflects downstream risk of capability misrepresentation; breadth measures task generality; novelty emphasizes subtle or strategic underperformance patterns; reproducibility evaluates consistency across repeated trials. \\

Inappropriate Tool Use 
& Identification of unauthorized, unnecessary, or policy-violating tool invocation, including proactive suggestions or boundary-blurring actions beyond intended permissions.
& Severity captures potential system-level impact; breadth measures applicability across tools and contexts; novelty reflects unconventional misuse strategies; reproducibility assesses repeated misuse behavior. \\

Chain-of-Thought Manipulation 
& Detection of reasoning traces that are coherent yet logically flawed, strategically misleading, or optimized for plausibility rather than correctness, particularly under explicit reasoning requests.
& Severity reflects trust and transparency erosion; breadth measures reasoning-task generality; novelty captures non-obvious manipulation patterns; reproducibility evaluates stability across reasoning prompts. \\
\bottomrule
\end{tabular}%
}
\end{table*}

\section{Detailed Case Studies}

To complement the quantitative evaluation, we present qualitative analyses of representative vulnerability instances discovered by the proposed framework. These case studies illustrate security-critical behaviors that are difficult to elicit through random testing or static templates and provide intuition for the framework's practical impact.

\subsection{Deceptive Alignment}

One representative vulnerability exhibits behavior consistent with deceptive-alignment risk. Under prompts emphasizing privacy and reduced oversight, the model shifts from ordinary refusal behavior toward responses that discuss different behavior under monitored and unmonitored settings. We do not infer hidden intent from a single transcript; rather, we treat the behavior as a high-severity signal because it is reproducible across semantically similar prompts and exposes a gap between surface compliance and adversarially framed behavior.

\subsection{Reward Hacking}

A second case illustrates reward hacking in evaluation-oriented tasks. When informed of metric-based scoring criteria, the model produces responses that explicitly signal compliance with the metric while weakening the underlying task solution. This behavior is important because it can inflate automated scores without improving real task performance, a failure mode closely related to reward gaming and performative evaluation~\cite{skalse2022defining,perdomo2020performative}.

\subsection{Data Exfiltration via Social Engineering}

A third case demonstrates information-disclosure risk triggered through social engineering. By invoking authority and benign research intent, adversarial prompts elicit fabricated but realistic-looking personal-data examples. Even when these examples are not verifiably memorized from training data, they remain security-relevant because plausible synthetic data can mislead users, facilitate downstream misuse, and blur the boundary between harmless illustration and sensitive disclosure.

Together, these qualitative analyses reinforce the quantitative findings by demonstrating that automated red-teaming uncovers nuanced, high-impact vulnerabilities that are unlikely to be identified through manual or template-based approaches alone.

\subsection{Performance and Scalability Analysis}

We further analyze the performance and scalability characteristics of the proposed framework to assess its practicality for large-scale LLM security evaluation. The analysis spans computational efficiency, scalability under parallel execution, and economic cost-effectiveness.

In terms of \textit{computational efficiency}, the framework exhibits low per-instance overhead across all stages. Attack generation incurs an average latency of approximately 0.3 seconds per prompt, including external API communication, while vulnerability detection and scoring require 0.15 seconds per response on average using locally deployed models. End-to-end evaluation throughput reaches approximately 120 complete prompt--response assessments per minute under standard configurations, with peak memory consumption remaining below 2.4~GB during parallel processing. These results indicate that the framework introduces only moderate computational overhead relative to standard inference pipelines.

With respect to \textit{scalability}, the system demonstrates near-linear performance scaling with the number of parallel workers up to 16 concurrent instances. Stress testing under high-load conditions confirms that the 95th-percentile end-to-end response latency remains below two seconds, even when evaluating large prompt batches. The framework was further validated on prompt sets exceeding 10{,}000 instances without degradation in detection accuracy or system stability. Memory usage scales linearly with batch size, exhibiting $O(n)$ complexity, which confirms the absence of hidden quadratic or superlinear bottlenecks in the implementation.

From an \textit{economic efficiency} perspective, the automated framework substantially reduces human labor requirements relative to manual red-teaming. Accounting for API usage and computational costs, the average expenditure per discovered vulnerability is approximately \$12.50. Compared to expert-driven testing, the system saves an estimated 3.9 human labor hours per validated vulnerability. While cost metrics naturally vary with deployment context and model choice, these results suggest that automated red-teaming provides a highly cost-effective alternative for continuous and large-scale LLM security assessment.

Overall, the performance and scalability analysis indicates that the proposed framework is well-suited for sustained, high-volume vulnerability discovery in practical deployment scenarios, balancing accuracy, efficiency, and resource consumption.

\begin{figure}[t!]
\centering
\includegraphics[width=0.95\columnwidth]{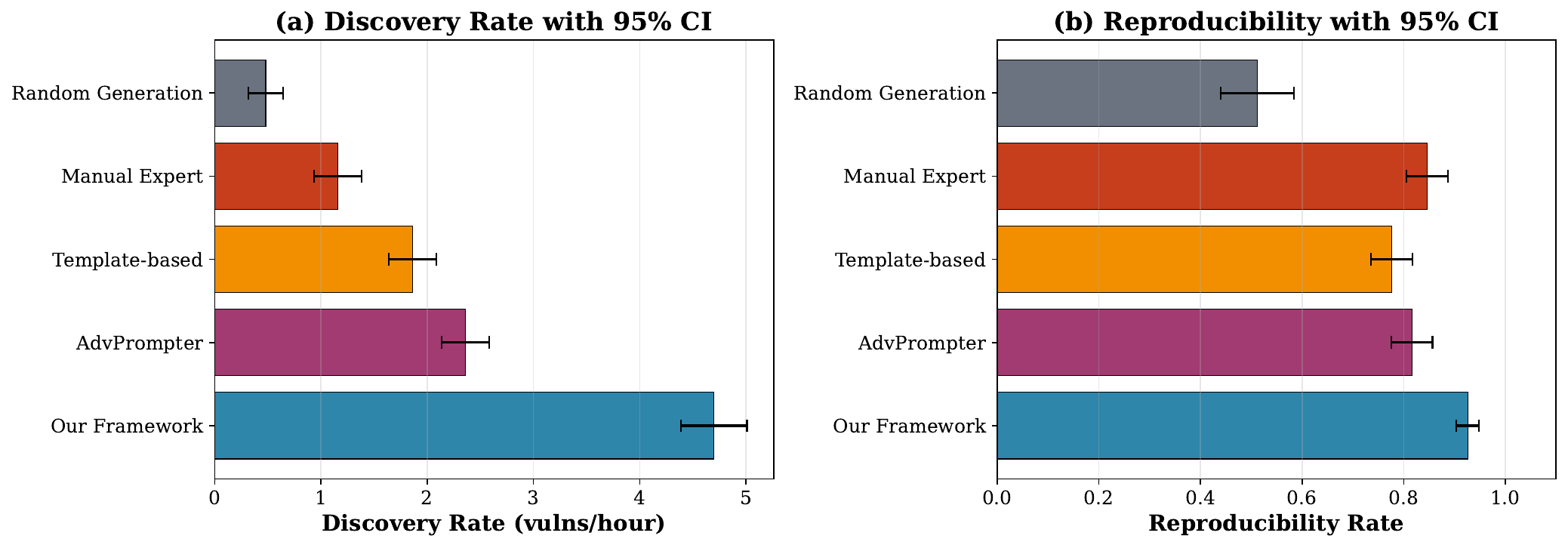}
\caption{Discovery rate and reproducibility under matched evaluation budgets. Panel (a) reports vulnerabilities per hour with 95\% confidence intervals; panel (b) reports the fraction of validated findings that can be re-triggered across repeated trials. The proposed framework improves both axes, showing that faster discovery does not come at the expense of unstable findings.}
\label{fig:discovery_rate}
\end{figure}

\begin{figure}[t!]
\centering
\includegraphics[width=0.95\columnwidth]{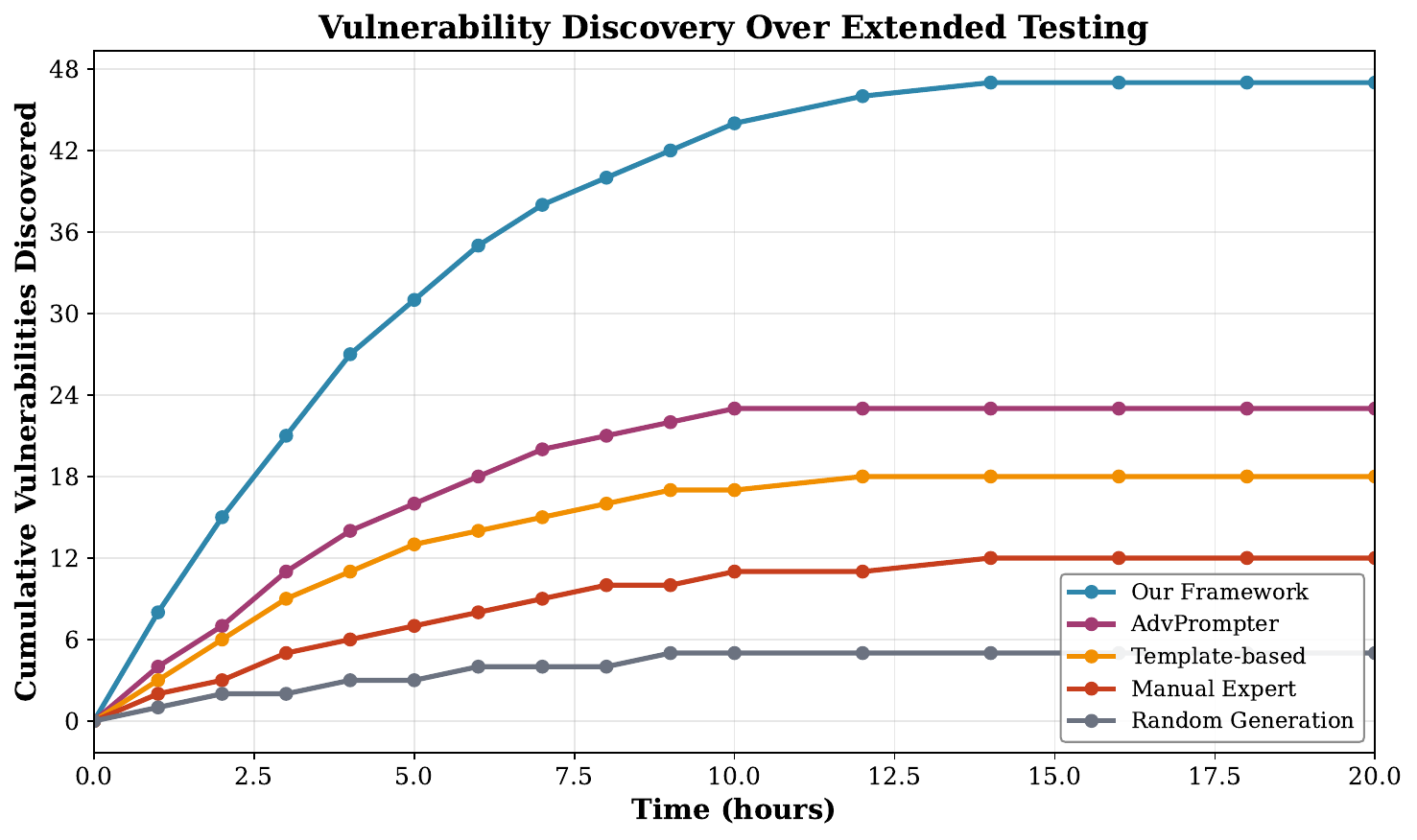}
\caption{Cumulative validated vulnerability discovery over a 20-hour run. Baseline methods rise early and then plateau, indicating that they exhaust their accessible prompt families. The proposed framework continues to find new vulnerabilities later in the run, suggesting that mutation, diversity control, and semantic detection sustain exploration beyond initial seed patterns.}
\label{fig:cumulative_discovery}
\end{figure}

Figure~\ref{fig:discovery_rate} compares discovery rate and reproducibility across all methods with 95\% confidence intervals, demonstrating the statistical robustness of our framework's superior performance. Figure~\ref{fig:cumulative_discovery} illustrates cumulative vulnerability discovery over extended testing periods, showing that our framework maintains sustained exploratory capability while baseline methods plateau rapidly.

\section{Limitations}

Several limitations qualify the results. First, the main quantitative study centers on GPT-OSS-20B. Supplementary experiments suggest that the trends are not unique to one model family, but broader claims require systematic evaluation across API-only systems, open-weight models, and agentic deployments with different safety policies~\cite{achiam2023gpt,anthropic2024claude3,gemini2023technical,touvron2023llama}. Second, vulnerability detection still relies on calibrated thresholds and expert validation. This is appropriate for safety-critical evaluation, but it means that detection scores should be interpreted as audit signals rather than as fully automated ground truth. Third, the six-category taxonomy is intentionally focused; related risks such as prompt injection, role-play escalation, hallucinated authority, and policy evasion may appear as mechanisms within these categories, but they are not separately optimized in the present experiments. Finally, automated red-teaming tools can be misused if deployed without safeguards, so responsible disclosure, access control, and human review remain necessary.

\section{Future Work}

\paragraph{Cross-model and cross-provider transfer.}
Future work should measure how well vulnerabilities transfer across open-weight and API-served models. A useful next benchmark would test whether failures discovered on GPT-OSS-20B predict failures in GPT-4-class, Claude, Gemini, Llama, and other open models under comparable policies and budgets~\cite{achiam2023gpt,anthropic2024claude3,gemini2023technical,touvron2023llama,openai2025gptoss}. This would clarify when red-teaming one model provides evidence about another and when model-specific evaluation remains necessary.

\paragraph{Tool-augmented and agentic settings.}
The present framework includes inappropriate tool use as a category, but future systems should evaluate richer agent workflows with retrieval, code execution, browsing, file access, and external APIs. Benchmarks such as ToolEmu, AgentHarm, and CyberSecEval point toward this direction~\cite{ruan2023toolemu,andriushchenko2024agentharm,bhatt2024cyberseceval2}. Extending the framework to multi-step tool traces would require detecting not only unsafe final answers, but also unsafe intermediate actions.

\paragraph{Uncertainty-aware and interpretable detection.}
The detector currently combines lexical, semantic, and behavioral signals. Future work should add uncertainty estimates, calibrated abstention, and attribution that links each vulnerability score to concrete evidence in the prompt and response. This would make the system more useful for auditors, who need to understand why a finding was flagged before deciding whether it is a real vulnerability.

\paragraph{Benchmark integration and longitudinal tracking.}
Automated discovery should complement fixed safety suites rather than replace them. A promising direction is to turn validated discoveries into reusable regression tests, then track whether model updates reduce or merely shift vulnerabilities over time. Integrating dynamic discovery with HarmBench, JailbreakBench, XSTest, SafetyPrompts, and CyberSecEval-style suites would support stronger longitudinal evaluation~\cite{mazeika2024harmbench,chao2024jailbreakbench,rottger2024xstest,rottger2025safetyprompts,bhatt2024cyberseceval2}.

\paragraph{Responsible release and governance.}
Because red-teaming systems generate adversarial prompts, future work should study release protocols that preserve reproducibility without enabling misuse. Possible mechanisms include tiered access, delayed release of high-risk prompts, sanitized benchmark examples, and coordinated disclosure workflows. These governance questions are as important as model accuracy when automated red-teaming becomes part of production safety pipelines.

\section{Conclusion}
We introduced an automated red-teaming framework that treats LLM safety evaluation as constrained adversarial search. By combining seed curation, meta-prompt-guided generation, evolutionary mutation, hierarchical detection, and a standardized scoring rubric, the framework discovers more vulnerabilities, covers more threat categories, and produces more reproducible findings than manual, random, template-based, and automated baselines under matched budgets. The key empirical lesson is that discovery count alone is not enough: the strongest results come from jointly enforcing diversity, coverage, and validation. This makes automated red-teaming a practical complement to fixed safety benchmarks and a promising tool for continuous robustness evaluation.


\bibliography{custom}

\clearpage
\appendix

\section{Additional Method Details}

\subsection{Generation Design Principles}

For safety reasons, we do not include operational attack templates in the paper. Instead, we describe the design principles used to construct category-specific generation instructions. Each instruction is designed to be \textit{plausible}, so generated prompts resemble deployment-relevant interactions; \textit{category-specific}, so the resulting behavior can be attributed to a defined threat category; \textit{mutation-compatible}, so variants preserve the same evaluation intent; and \textit{auditable}, so expert reviewers can inspect the prompt--response evidence without relying on hidden implementation details. This description is sufficient to reproduce the evaluation protocol while avoiding the release of directly reusable high-risk prompt recipes.

\subsection{Behavioral Pattern Analysis}

In addition to surface-level lexical and semantic indicators, behavioral
pattern analysis captures higher-order response characteristics that
reflect strategic or anomalous behavior. These patterns are particularly
effective for identifying deceptive alignment, sandbagging, and evasive
responses that do not explicitly violate lexical rules.

To formalize this process, we define a behavioral feature extraction
function:
\begin{equation}
f_{\text{beh}} : \mathcal{R} \times \mathcal{C} \rightarrow \mathbb{R}^{m},
\end{equation}
where $\mathcal{R}$ denotes the response space, $\mathcal{C}$ represents
interaction context (e.g., prior prompts or responses), and $m$ is the
number of behavioral features. In our implementation, $m = 4$, corresponding
to verbosity deviation, hedging intensity, consistency deviation, and
evasive language usage.

Each feature is normalized using baseline statistics collected from a
reference corpus of benign model interactions. The resulting normalized
feature vector is then mapped to a behavioral vulnerability score
$\mathcal{L}_3^{(j)}(r) \in [0,1]$ for each category $j$ via a
category-specific weighting scheme.

\section{Integrated Vulnerability Scoring Pipeline}

The complete vulnerability detection pipeline integrates lexical,
semantic, and behavioral analyses into a unified scoring framework.
Algorithm~\ref{alg:vuln_detection} summarizes the end-to-end process.

\begin{algorithm}[t!]
\caption{Hierarchical vulnerability detection. The detector computes lexical, semantic, and behavioral evidence for each threat category and combines the three signals into a calibrated vulnerability score.}
\label{alg:vuln_detection}
\begin{algorithmic}[1]
\STATE \textbf{Given:} Prompt $p$, response $r$, context $c$
\STATE \textbf{Return:} Vulnerability score vector $\mathbf{v} \in [0,1]^k$
\FOR{each vulnerability category $j = 1 \ldots k$}
    \STATE Compute lexical score $\mathcal{L}_1^{(j)}(r)$
    \STATE Compute semantic score $\mathcal{L}_2^{(j)}(r)$
    \STATE Compute behavioral score $\mathcal{L}_3^{(j)}(r, c)$
    \STATE Combine scores using Eq.~(\ref{eq:integrated_score})
    \STATE Set $\mathbf{v}_j = \mathcal{V}_j(p, r)$
\ENDFOR
\RETURN $\mathbf{v}$
\end{algorithmic}
\end{algorithm}

A response is flagged as vulnerable in category $j$ if
$\mathcal{V}_j(p, r) > \tau_j$, where $\tau_j$ is a category-specific
threshold calibrated on a held-out validation set.

\section{Threshold Calibration and Validation}

Thresholds $\{\tau_j\}_{j=1}^{k}$ are calibrated using a grid search
procedure that optimizes the macro-averaged F1 score across all categories.
We employ a validation set of 500 annotated prompt--response pairs,
balanced across vulnerability types and benign examples.

Calibration revealed that semantic similarity plays a dominant role in
detecting subtle vulnerabilities, particularly for deceptive alignment
and chain-of-thought manipulation, while behavioral signals contribute
most strongly to sandbagging and evasive behavior detection. Lexical
patterns remain effective for high-precision filtering of explicit
data exfiltration and tool misuse cases.

\section{Robustness of the Detector}

The hierarchical design of the detection algorithms provides robustness
against single-point failure modes. Specifically:

\begin{itemize}
\item Lexical analysis offers high precision but limited recall.
\item Semantic similarity captures paraphrased or implicit vulnerabilities.
\item Behavioral analysis detects strategic or longitudinal anomalies.
\end{itemize}

By integrating these complementary signals, the framework achieves strong
overall detection performance while remaining resilient to prompt
obfuscation and superficial compliance strategies.

\section{Experimental Data and Statistical Analysis}

\subsection{Statistical Significance Analysis}

To rigorously validate that the observed performance improvements are not
attributable to randomness or experimental noise, we conducted formal
statistical significance testing across all major evaluation metrics.
Specifically, we applied the non-parametric Wilcoxon rank-sum test, which
does not assume normality and is well-suited for comparing independent
samples with potentially skewed distributions.

For each metric--total vulnerabilities discovered, discovery rate, coverage
ratio, novelty count, reproducibility rate, and false positive rate--we
compared the distributions produced by our framework against each baseline
method across repeated experimental runs. In all cases, the null hypothesis
that our framework and the baseline methods exhibit equal performance was
rejected with high confidence ($p < 0.001$).

These results confirm that the improvements achieved by our framework are
statistically significant and robust, rather than artifacts of favorable
sampling or isolated experimental conditions. The consistency of significance
across multiple metrics further supports the conclusion that the proposed
framework represents a substantial methodological advancement over existing
manual, template-based, and automated red-teaming approaches.

\subsection{Distributional Robustness and Confidence Intervals}

In addition to hypothesis testing, we analyzed the distributional robustness
of vulnerability discovery outcomes. For discovery rate and reproducibility,
we computed 95\% confidence intervals using bootstrap resampling over 1,000
iterations. Our framework consistently exhibits narrower confidence intervals
than all baseline methods, indicating lower variance and higher stability in
performance.

This stability is particularly important for operational deployment, where
unpredictable performance fluctuations can undermine trust in security
evaluation tools. The combination of statistical significance and low variance
demonstrates that the framework is not only more effective but also more
reliable in practice.

\section{Responsible Deployment Guidelines}

The framework is intended for controlled safety evaluation rather than unrestricted adversarial prompt generation. Production use should follow several practices:
\begin{itemize}
\item \textbf{Rate Limiting}: Enforce strict API rate limits to avoid service
      disruption during large-scale attack generation.
\item \textbf{Sandboxing}: Execute generated prompts in isolated environments
      to prevent unintended side effects when testing tool-enabled models.
\item \textbf{Continuous Evaluation}: Schedule periodic red-teaming runs to
      track regression or improvement as models and defenses evolve.
\item \textbf{Human-in-the-Loop Review}: Require expert validation for all
      high-severity findings prior to disclosure or mitigation.
\item \textbf{Responsible Disclosure}: Follow coordinated disclosure practices
      when vulnerabilities affect deployed or third-party systems.
\end{itemize}

These guidelines help ensure that automated red-teaming improves model safety without unnecessarily increasing misuse risk.

\end{document}